\documentclass[12pt]{article}

\renewcommand{\theequation}{\thesection.\arabic{equation}}

\title{Spikes in Cosmic Crystallography}
\author{G.I. Gomero\thanks{E-mail: german@cbpf.br} , \  
A.F.F. Teixeira\thanks{E-mail: teixeira@cbpf.br} , \ 
M.J. Rebou\c{c}as\thanks{E-mail: reboucas@cbpf.br} \ 
\& A. Bernui\thanks{E-mail: bernui@fc-uni.edu.pe.
Permanent address: Facultad de Ciencias,
Universidad Nacional de Ingenier\'{\i}a, 
Apartado 31-139, \  Lima 31 -- Peru.} \\ 
\\ 
Centro Brasileiro de Pesquisas F\'\i sicas, \\
Rua Dr. Xavier Sigaud 150 \\
22290-180 Rio de Janeiro -- RJ, Brazil
}

\begin{document}
\date{}
\maketitle

\begin{abstract}
If the universe is multiply connected and small the sky 
shows multiple images of cosmic objects, correlated by the
covering group of the 3-manifold used to model it.
These correlations were originally thought to manifest 
as spikes in pair separation histograms (PSH) built from 
suitable catalogues. 
Using probability theory we derive an expression for the 
expected pair separation histogram (EPSH) in a rather 
general topological-geometrical-observational setting. 
As a major consequence we show that the spikes of 
topological origin in PSH's 
are due to translations, whereas other isometries manifest  
as tiny deformations of the PSH corresponding to the 
simply connected case. This result holds for all 
Robertson-Walker spacetimes and gives rise to two basic
corollaries: (i) that PSH's of Euclidean manifolds that have 
the same translations in their covering groups exhibit 
identical spike spectra of topological origin, making clear 
that even if the universe is f\/lat the topological spikes 
alone are not suf\/f\/icient for determining its topology; 
and (ii) that PSH's of hyperbolic 3-manifolds exhibit 
no spikes of topological origin. These corollaries ensure 
that cosmic crystallography, as originally formulated, is 
not a conclusive method for unveiling  the shape of the 
universe. We also present a method that reduces the statistical
f\/luctuations in PSH's built from simulated catalogues.
\end{abstract}

\section{Introduction}
\label{intro}
\setcounter{equation}{0}

Current observational data favour Friedmann-Lema\^{\i}tre (FL)
cosmological models as approximate descriptions of our universe at
least since the recombination time. These descriptions are, however,
only local and do not f\/ix the global shape of our universe. 
Despite the inf\/initely many possibilities for
its global topology, it is often assumed that spacetime is
simply connected leaving aside the hypothesis, very rich in
observational and physical consequences, that the universe may
be multiply connected, and compact even if it has zero or
negative constant curvature. Since the hypothesis that
our universe has a non-trivial topology has not been excluded,
it is worthwhile testing it (see~\cite{CosmicTop}~--~\cite{ZelNov83} 
and references therein). 

The most immediate consequence of the hypothesis of multiply-%
connectedness of our universe is the possibility of observing
multiple images of cosmic objects, such as galaxies, quasars, and 
the like. Thus, for example, consider the available catalogues of 
quasars with redshifts ranging up to $z \approx 4$ which, in
the Einstein-de Sitter cosmological model, corresponds to a
comoving distance $d \approx 3300 \, h^{-1} Mpc$ from us ($h$
is the Hubble constant in units of $100 \, km \, s^{-1} 
Mpc^{-1}$). Then, roughly speaking, if our universe is \emph{small} 
in the sense that it has closed geodesics of length less than 
$2 d$, some of the observed quasars may actually be images of 
the same cosmic object.%
\footnote{Note that we are not considering problems which 
arise from the possibly short lifetime of quasars. Actually, 
this is irrelevant for the point we want to illustrate
with this example.}

More generally, in considering discrete astrophysical sources,
the \emph{observable universe} can be viewed as that part of 
the universal covering manifold $\widetilde{M}(t_0)\,$ of the 
$t=t_0$ space-like section $M(t_0)$ of spacetime, causally
connected to an image of our position since the moment of 
matter-radiation decoupling (here $t_0$ denotes present time) 
while, given a catalogue of cosmic sources, the 
\emph{observed universe} is that part of the observable 
universe which contains all the sources listed in the catalogue. 
So, for instance, using quasars as cosmic sources the observed 
universe for a catalogue covering the entire sky is a ball with 
radius approximately half the radius of the observable universe 
(in the Einstein-de Sitter model). If the universe $M(t_0)$ is 
small enough in the above-specif\/ied sense, then there may be
\emph{copies} of some cosmic objects in the observed universe,
and an important goal in observational cosmic topology is to
develop methods to determine whether these copies exist.

Direct searching for multiple images of cosmic objects is not a
simple problem. Indeed, due to the f\/initeness of the speed of
light two images of a given object at dif\/ferent distances
correspond to dif\/ferent epochs of its life. Moreover, in 
general the two images are seen from dif\/ferent directions.
So one ought to be able to f\/ind out whether two images
correspond to dif\/ferent objects, or correspond to the same
object seen at two dif\/ferent stages of its evolution and at
two dif\/ferent orientations. The problem becomes even more
involved when one takes into account that observational and 
selection ef\/fects may also be dif\/ferent for these distinct 
images.

One way to handle these dif\/f\/iculties is to use suitable
statistical analysis applied to catalogues. Cosmic crystallography%
~\cite{LeLaLu} is a promising statistical method which looks for 
distance correlations between cosmic sources using pair separation 
histograms (PSH), i.e. graphs of the number of pairs of sources 
versus the squared distance between them. These correlations are 
expected to arise from the isometries of the covering group of 
$M(t_0)$ which give rise to the (observed) multiple images, and 
have been claimed to manifest as sharp peaks~\cite{LeLaLu}, also 
called spikes. Moreover, the positions and relative amplitudes 
of these spikes have also been thought to be f\/ingerprints 
of the shape of the universe (see, however, the references%
~\cite{LeLuUz}~--~\cite{FagGaus}, and also
sections~\ref{fres} and~\ref{concl} of the present paper).

It should be emphasized that just by examining a single PSH  
one cannot at all decide whether a particular spike is of 
topological origin or simply arises from statistical f\/luctuations. 
Actually, depending on the accuracy of the simulation one can obtain 
PSH's with hundreds of sharp peaks of purely statistical origin among just a 
few spikes of topological nature. Thus, it is of indisputable importance 
to perform a theoretical statistical analysis of the distance 
correlations in the PSH's at least to have a criterion for
revealing the ultimate origin of the spikes that arise 
in the PSH's and pave the way for further ref\/inements of the
crystallographic method.

In this work, by using the probability theory we derive the 
expression~(\ref{EPSH2}) for the expected pair separation histogram 
(EPSH) of comparable catalogues with the same number of sources and 
corresponding to any complete 3-manifold of constant curvature.
The EPSH, which is essentially a PSH from which the statistical noise 
has been withdrawn, is derived in a very general topological-geometrical%
-observational setting. 
It turns out that the EPSH built from a multiply connected manifold is 
an EPSH in which the contributions arising from the correlated images 
were withdrawn, plus a term that consists of 
a sum of individual contributions from each covering isometry.

{}From the EPSH~(\ref{EPSH2}) we extract its major consequence, 
namely that the sharp peaks (or spikes) of topological nature in 
individual pair separation histograms are due to Clif\/ford 
translations, whereas all other isometries manifest as tiny 
deformations of the PSH of the corresponding universal covering 
manifold.
This relevant consequence holds for all Robertson-Walker (RW) spacetimes 
and in turn gives rise to two others: (i) that Euclidean distinct manifolds 
which have the same translations in their covering groups present 
the same spike spectra of topological nature. So, the set of topological
spikes (their positions and relative amplitudes) in the PSH alone is not 
suf\/f\/icient for distinguishing these compact f\/lat manifolds, making
clear that even if the universe is f\/lat ($\Omega_{tot}=1$) the spike 
spectrum is not enough for determining its global shape; and
(ii) that individual PSH's corresponding to hyperbolic 3-manifolds 
exhibit no spikes of topological origin, since there are no Clif\/ford 
translations in the hyperbolic geometry.
These two corollaries ensure that cosmic crystallography, as originally 
formulated, is not a conclusive method for unveiling the shape of the 
universe.
 
As a way to reduce the statistical f\/luctuations in individual 
PSH's so as to unveil the contributions of non-translational 
isometries to the topological signature, we introduce the 
mean pair separation histogram (MPSH) and show that 
it is a suitable approximation for the EPSH. 
Moreover, we emphasize that the use of MPSH's is restricted 
to simulated catalogues due to the unsurmountable practical 
dif\/f\/iculties in constructing several comparable catalogues 
of real sources.
 
The lack of spikes of topological origin in PSH's of multiply connected 
hyperbolic 3-manifolds has also been found in histograms for the specif\/ic 
cases of Weeks~\cite{LeLuUz} and one of the Best~\cite{FagGaus} manifolds, 
making apparent that, within the degree of accuracy of the corresponding 
plots, the above corollary (ii) holds for these specific cases.
Concerning these references, we also discuss the connection between 
ours and their~\cite{LeLuUz,FagGaus} results. Further, we point out 
the limitation of the set of conclusions one can withdraw from 
such graphs by using, e.g., the specif\/ic PSH shown in f\/ig.~1 of%
~\cite{Fagundes-Gausmann}. The possible origins for the 
spikes in PSH's are also discussed, and it is shown that one can 
distinguish between statistical sharp peaks (noise) and topological 
spikes only through the use of the rather general results obtained 
in this article.

The plan of this paper is as follows. In the next section we 
describe what a catalogue of cosmic sources is in the context of 
Robertson-Walker (RW) spacetimes, and introduce some relevant 
def\/initions to set our framework and make our paper as accurate 
and self-contained as possible.
In the third section we describe how to construct a PSH from a
given catalogue (either real or simulated), and discuss
qualitatively how distance correlations arise in multiply 
connected RW universes. In the fourth section we f\/irst discuss the
concept of \emph{expected} pair separation histogram (EPSH) and
derive its explicit expression in a very general 
topological-geometrical-observational setting.
In the f\/ifth section we use the expression for the EPSH obtained 
in section~\ref{pshexp} to derive its most general consequence, 
namely that spikes of topological origin in PSH's are due to 
\emph{translations} alone. We proceed by analyzing how this general 
result af\/fects PSH's built from Euclidean and hyperbolic manifolds,
and derive two relevant consequences regarding these classes of manifolds. 
We also present in that section the MPSH as a simple approach 
aiming at reducing the statistical f\/luctuations in PSH's 
so that the contributions from non-translational isometries 
become apparent. 
In section~\ref{concl} we summarize our conclusions,
brief\/ly indicate possible approaches for further 
investigations, and f\/inally discuss the connection between 
ours and the results those reported in~\cite{LeLuUz} 
and~\cite{FagGaus}. 

\section{Catalogues in multiply connected RW spacetimes}
\label{catalogs}
\setcounter{equation}{0}

If the universe is multiply connected and one can form 
catalogues of cosmic sources with multiple images, the problem 
of identifying its shape can be reduced to that of designing 
suitable methods for extracting the underlying topological 
information from these catalogues. In this section we describe
what a catalogue of cosmic sources is in the context of 
RW spacetimes, and discuss under what conditions 
catalogues present multiple images. We f\/irst brief\/ly 
review some basic properties of locally homogeneous and isotropic
cosmological models, then we formalise the practical process of 
construction of catalogues of discrete astrophysical sources, 
and f\/inally we specify the conditions for the existence of
multiple images in a catalogue.

\vspace{6mm}
\noindent \textbf{\large Locally homogeneous and isotropic cosmological
             models}
\vspace{3mm}

The spacetime arena for a FL cosmological model is a 
4-dimensional manifold endowed with a RW metric which 
can be written locally as
\begin{equation}
\label{RW-metric}
ds^2 = dt^2 - a^2(t)\, d \sigma^2 \; ,
\end{equation}
where $t$ is a cosmic time, $d\sigma$ is a standard 3-dimensional
hyperbolic, Euclidean or spherical metric, and $a(t)$ is the
scale factor that carries the unit of length. It is usually
assumed that the $t=const$ spatial sections of a RW spacetime
are one of the following simply connected spaces: hyperbolic ($H^3$), 
Euclidean ($E^3$), or the 3-sphere ($S^3$), depending on the local
curvature computed from $d \sigma$. Cosmic topology arises when
we relax the hypothesis of simply-connectedness and consider that
the $t=const$ spatial sections may also be any complete multiply 
connected 3-manifold of constant curvature (see, for example,
\cite{Wolf,Ellis}).

In this work  we shall consider spacetimes of the form 
$I \times M$, with $I$ a (possibly inf\/inite) open interval of
the real line, and $M$ a complete constant curvature 3-manifold, 
either simply or multiply connected. Actually, a RW spacetime is 
a warped product $I \times_a M$ (see, e.g.,~\cite{O'Neill} for 
more details) in that, for any instant $t \in I$, the metric in 
$M$ is $d\sigma (t) = a(t) d\sigma$. The manifold $M$ equipped 
with the metric $d\sigma (t)$ is denoted by $M(t)$, and
is called \emph{comoving space} at time $t$. 
So, \emph{comoving geodesics} and \emph{comoving distances}
at some time $t$ mean, respectively,  geodesics of $M(t)$ and
distances between points on $M(t)$. 
Throughout this article we will omit the time dependence
of $M(t)$ and $\widetilde{M}(t)$ whenever these manifolds
are endowed with the standard metric $d\sigma\,$. 

Before proceeding to the discussion of the notion of catalogues
we recall that in a FL cosmological model the energy-matter content
and Einstein's f\/ield equations determine the local curvature of 
the spatial sections and the scale factor $a(t)$. 
Thus in this process of cosmological modelling within the framework 
of general relativity the introduction of particular values for 
cosmological parameters ($H_0$, $\Omega_m$, $\Omega_{\Lambda}$, 
$q_0$, and so forth) restricts both the 
locally homogeneous-and-isotropic 3-geometry, and  
the scalar factor $a(t)$. However, the concepts and results 
we shall introduce and derive in this work hold regardless 
of the particular 3-geometry and of the form of $a(t)$ provided 
that $a(t)$ is a monotonically increasing function, at least 
since the recombination time. 
So, there is def\/initely no need to introduce any particular 
values for the cosmological parameters unless one intends 
to examine the consequences of a specif\/ic class of RW models, 
which for the sake of generality we do not aim at in the present 
article.

\vspace{6mm}
\noindent \textbf{\large Constructing catalogues}
\vspace{3mm}

To formalise the concept of catalogue of cosmic sources, let us 
assume that we know the scale factor $a(t)$, and that it is a
monotonically increasing function. We shall also assume that all
cosmic objects of our interest (also referred to simply as
objects) are pointlike and have long lifetimes so that none
was born or dead within the time interval given by $I$. Moreover,
we shall also assume that all objects are comoving, so that their
worldlines have constant spatial coordinates. Although unrealistic,
these assumptions were used implicitly in~\cite{LeLaLu} and%
~\cite{Fagundes-Gausmann} and are very useful to study the 
observational consequences of a non-trivial topology for the
universe.

It should be noted that in the process of construction of 
catalogues we shall describe below it is assumed that a 
particular type of sources (clusters of galaxies, quasars, etc) 
or some combination of them (quasars and BL Lac objects, say) is 
chosen from the outset. 
This approach does not coincide with the exact manner the 
astronomers build catalogues, in that usually they simply 
record any sources within their range of interest and (or) 
observational limitations. However, the model of constructing 
catalogues we shall present relies on the fact that any catalogue
of a specif\/ic type of sources ultimately is a selection of 
sources of that type from the hypothetical complete set of sources 
which can in principle be observed.

Since we are assuming that all objects are comoving their spatial 
coordinates are constant. So, the set of all the objects in $M$ is 
given by a list of their present comoving positions, and from this 
list one can def\/ine a map
\begin{eqnarray} 
\qquad \qquad \mu : M(t_0) & \rightarrow & \{1,0\} \nonumber \\
p \:  \mapsto \mu(p) &=& \left\{ \begin{array}
{r@{\quad}l}
1 & \mbox{if there is an object at $p\;$,} \\
0 & \mbox{otherwise}\;.   
\end{array} \right.
\end{eqnarray}

The set of objects in $M(t_0)$ is thus $\mu^{-1}(1)$.
This is a discrete set in $M(t_0)$ without accumulation points.
Actually, from any  map $\mu : M(t_0) \rightarrow \{1,0\}$ such
that $\mu^{-1}(1)$ is  a discrete set without accumulation points,
one may def\/ine a set of objects, namely the set $\mu^{-1}(1)$.
We will further assume in this work that the set of objects is a
representative sample of some well-behaved distribution law in 
$M(t_0)$. For our purposes in this article a distribution is 
well-behaved if it gives rise to samples of points which are 
not concentrated in small regions of $M(t_0)$.

Let $\pi: \widetilde{M}(t_0) \rightarrow M(t_0)$ be the
universal covering projection of $M(t_0)$ and $p$ be an
object, that is $p \in \mu^{-1}(1)$. The set $\pi^{-1}(p)$
is the collection of \emph{copies} of $p$ on $\widetilde{M}(t_0)$.
We will refer to these copies as \emph{topological images} or 
simply as images of the object $p$, thus the map $\tilde{\mu}$ 
def\/ined by the commutative diagram 
\\
\setlength{\unitlength}{1.5cm}
\begin{picture}(3,3)
\put(4.15,2){$\widetilde{M}(t_0)$}
\put(4.5,1.85){\vector(0,-1){1}}
\put(4.9,2){\vector(1,-1){1.15}}
\put(4.15,0.5){$M(t_0)$}
\put(4.9,0.57){\vector(1,0){1}}
\put(6,0.5){$\{1,0\}$}
\put(4.25,1.3){$\pi$}
\put(5.3,0.35){$\mu$}
\put(5.6,1.5){$\tilde{\mu}$}
\end{picture} 
\\
gives the set of all images on the universal covering manifold 
$\widetilde{M}(t_0)$.%
\footnote{It has occasionally been used in the 
literature a misleading terminology in which the topological 
images are classif\/ied as real and ghosts. In most cases the 
expression `real image' refers to the nearest 
topological image of a given object,
while the  expression `ghost image' refers to any other image 
of the same object. There are also cases where `real images' 
has been used to refer to the topological images which lie 
inside a fundamental polyhedron (FP), while `ghost images' refers 
to the images outside the FP. This latter classif\/ication of the 
images depends on the choice of the FP which is not at all unique. 
Actually, these two usages for the expressions `real image' and 
`ghost images' are compatible only if the FP is the Dirichlet 
domain with centre at an image of the observer~\cite{Images}. 
In both cases the terminology is misleading also because it 
may suggest that either the nearest images or the images inside a FP 
are somehow special. And yet there is no physical or geometrical 
property which supports this distinction. Besides, this terminology 
is unnecessary since real objects are represented by points 
which lie on the manifold $M$, whereas the points on the universal 
covering $\widetilde{M}$ can represent only (topological) images of 
these objects --- no image is more \emph{real} than the other,  
they are simply (topological) images.}  
Indeed, the images of the objects in 
$M(t_0)$ are the elements of the set $\tilde{\mu}^{-1}(1)$.

Now suppose we perform a full sky coverage survey for the objects 
up to a redshift cutof\/f $z_{max}$. Since we are assuming that 
we know the metric $a(t)d\sigma$, we can compute the
distance $R$ corresponding to this redshift cutof\/f and so
determine the observed universe corresponding to this survey.
The ball $\mathcal{U} \subset \widetilde{M}(t_0)$ with radius
$R$ and centred at an image of our position is a representation
of this observed universe. 
The f\/inite set $\mathcal{O} = \tilde{\mu}^{-1}(1)\cap \mathcal{U}$ 
is the set of \emph{observable images} since it contains all the 
images which can in principle be observed up to a distance $R$ from 
one image of our position.

The set of \emph{observed images} or \emph{catalogue} is a subset 
$\mathcal{C} \subset \mathcal{O}$, since by several observational 
limitations one can hardly record all the images present in the 
observed universe. Our observational limitations can be formulated 
as \emph{selection rules} which describe how the subset $\mathcal{C}$ 
arises from $\mathcal{O}$. These selection rules, together with the 
distribution law which the objects in $M$ obey, will be referred 
to as \emph{construction rules} for the catalogue $\mathcal{C}$.
A good example of construction rules appears in the 
simulated catalogue constructed in ref.~\cite{LeLaLu} where an 
uniform distribution of points in a 3-torus is assumed together 
with a selection rule which dictates how one obtains a catalogue 
$\mathcal{C}$ from the set of images in an observed universe 
$\mathcal{U}$ subjected to (def\/ined by) the redshift cutof\/f 
$z_{max}=0.26$. In that example, to mimic the obscuration ef\/fect 
by the galactic plane, they have taken as selection rule that only 
the images inside a double cone of aperture $120^\circ$ are observed 
or considered. However, in more involved simulations, one can 
certainly take other selection rules such as, e.g., luminosity 
threshold, f\/inite lifetime and the obscuration by the line of 
sight, and (or) a combination of them.

\begin{sloppypar}
Throughout this work we shall assume that catalogues obey well-% 
def\/ined construction rules, and we shall say that two catalogues 
are \emph{comparable} when they are def\/ined by the same 
construction rules; even if they have a signif\/icantly 
dif\/ferent number of sources and correspond to possibly 
dif\/ferent (topologically) 3-manifolds compatible with a given 
3-geometry. 
According to this def\/inition two comparable catalogues 
must correspond to a given RW geometry~(\ref{RW-metric}) with 
obviously a well-def\/ined scale factor, and the same underlying 
redshift cutof\/f. In other words, 
comparable catalogues correspond to a precise set of cosmological 
parameters of an observed universe, plus a f\/ixed redshift 
cutof\/f. The main motivation for formalising in this way the 
concept of comparable catalogues comes from the fact that in 
the cosmic crystallographic method we are often interested in 
comparing PSH's from simulated catalogues against PSH's from real 
catalogues. And real catalogues are limited by a redshift cutof\/f 
that is converted into distance through an \emph{ad hoc} choice of 
a RW geometry. So, to build simulated catalogues comparable to a
specif\/ic real catalogue, for example, one has to begin with 
the precise RW geometry that transforms redshifts into distances 
in the real catalogues, and use the same redshift cutof\/f of the
underlying real catalogue. 

\end{sloppypar}

Finally, it should be noticed that the above def\/inition for a catalogue
f\/its in with the two basic types of catalogues one usually f\/inds
in practice, namely \emph{real catalogues} (arising from 
observations) and \emph{simulated catalogues}, which are generated
under well-def\/ined assumptions that are posed to mimic some 
observational limitations and (or) to account for simplif\/ying 
hypotheses.

\vspace{6mm}
\noindent \textbf{\large Catalogues with multiple images}
\vspace{3mm}

If $M$ is simply connected then $M$ and $\widetilde{M}$ are the
same, and so there is exactly one image for each object. If $M$
is multiply connected then each object has several images (actually an
inf\/inite number of images in the cases of zero and negative
curvature). Suppose that $M$ is multiply connected, and let $P
\subset \widetilde{M}(t_0)$ be a fundamental domain of $M(t_0)$.
$P$ can always be chosen in such a way that $\tilde{\mu}^{-1}(1)
\cap \partial P = \emptyset$, where $\partial P$ is the boundary
of $P$. If we consider the universal covering
$\widetilde{M}(t_0)$ tessellated by $P$, then clearly
$\tilde{\mu}^{-1}(1)$ presents all the periodicities due to the
covering group $\Gamma$, in the sense that in each copy $gP$ of
the fundamental domain ($\,g \in \Gamma\,$) there is the same
distribution of images as in $P$.

To be able to guarantee the existence of multiple 
images in a catalog $\mathcal{C}$, we shall need the concept of a
\emph{deep enough} survey, which is a survey
whose corresponding observed universe $\mathcal{U}$ has the
property that for some fundamental polyhedron $P$, there are faces
$F$ and $F'$, identif\/ied by an isometry $g \in \Gamma$, and such
that some portions $E \subset F$ and $g(E) \subset F'$ are in the
interior of $\mathcal{U}$. In particular when $M$ is compact with 
some fundamental polyhedron lying inside the observed universe
$\mathcal{U}$, then this observed universe corresponds to a deep 
enough survey. To f\/ind out whether a full sky coverage survey is 
deep enough, in practice, all one needs to do is to determine the 
closest image of our position and using the metric $a(t)d \sigma$ 
compute the redshift $z_{thr}$ corresponding to half of that
distance. Any full sky coverage survey with redshift 
$z > z_{thr}$ is said to be a deep enough survey.%
\footnote{As a matter of fact $z_{thr}$ is the redshift corresponding
to the radius of the inscribed ball in the Dirichlet polyhedron of 
$M$ centred in an image of our position~\cite{Images}.}

When $M$ is multiply connected and the survey is deep enough the
set of observable topological images $\mathcal{O}$ contains multiple 
images of some cosmic objects. If, in addition, our observational 
capabilities allow the presence of multiple images in 
$\mathcal{C}$, then the catalogue has information on the 
periodicities due to the covering group $\Gamma$, and so
about the manifold $M$. Every pair of images of one object is 
related by an isometry of $\Gamma$. These pairs of images 
have been called \emph{gg-pairs}~\cite{LeLaLu}, however  
when referring to them collectively we shall use the term 
$\Gamma$-pairs, reserving the name $g$-pair for any pair related
by a specif\/ic isometry $g \in \Gamma$. The $\Gamma$-pairs in
$\mathcal{C}$ give rise to correlations in the positions of the
observed images. The main goal of any statistical approach to
cosmic topology based on discrete sources is to develop methods
to reveal these correlations. Cosmic crystallography is one such 
method, and uses PSH's to obtain the \emph{distance correlations}
which arise from these correlations in positions.

It should be stressed that there are two independent conditions
that must be satisf\/ied to have multiple images in a catalogue 
$\mathcal{C}$. Firstly, the survey has to be deep enough, so that 
in the observed universe $\mathcal{U}$ there must exist multiple 
observable images of cosmic objects.
Secondly, the selection rules, which dictate how one obtains a 
catalogue $\mathcal{C}$ from the observed universe $\mathcal{U}$,
must not be so restrictive as to rule out the possible multiple
images $\mathcal{C}$.
Clearly if the survey is not deep enough there is no chance of
having multiple images in $\mathcal{C}$, regardless of the
quality of the observations. On the other hand, even when the 
survey is deep enough, if the selection rules are too strict 
they may reduce the multiple images in $\mathcal{C}$ to a level 
that the detection of topology becomes impossible.

Finally, it should be noticed that when $M$ is simply connected, 
to any cosmic object corresponds just one image, so in this case 
there is an one-to-one correspondence between images and objects, 
and we can simply identify them as the same entity. Since we do 
not know a priori whether our universe is simply or multiply 
connected, we do not know if we are recording objects or just 
images in real catalogues, hence we say that a catalogue is formed 
by \emph{cosmic sources}.

\section{Pair separation histograms}
\label{histog}
\setcounter{equation}{0}

The purpose of this section is two-fold. F\/irst we shall
give a brief description of what a PSH is and how to construct
it. Then we shall describe qualitatively how distance
correlations arise in a particular multiply connected universe, 
motivating therefore the statistical analysis we will perform in
the next section.

To build a PSH we simply evaluate a suitable one-to-one function
$f$ of the distance $r$ between the cosmic sources of every pair 
{}from a given catalogue $\mathcal{C}$, and then count the number
of pairs for which these values $f(r)$ lie within certain
subintervals. These subintervals must form a partition of
the interval $(0,f(2R)]$, where $R$ is the distance from us
to the most distant source in the catalogue. Usually all the
subintervals are taken to be of equal length. The PSH is
just a plot of this counting. Actually, what we shall call
a PSH is a normalized version of this plot. The function
$f$ is usually taken to be the square function, whereas for very
deep catalogues it might be convenient to try some hyperbolic
function if we are, for example, dealing with open FLRW models. 
In line with the usage in the literature and to be specif\/ic, 
in what follows we will take $f$ to be the square function.
However, it should be emphasized that the results we obtain
here and in the next section hold regardless of this 
particular choice.

A formal description of the above procedure is as follows.
Given a catalogue $\mathcal{C}$ of cosmic sources we denote
by $\eta(s)$ the number of pairs of sources whose squared
separation is $s$. Formally, this is given by the function
\begin{eqnarray*}
\eta : (0,4R^2] & \rightarrow & [0,\infty) \\
s & \mapsto & \frac{1}{2} \, \mbox{Card}(\Delta^{-1}(s))\;,
\end{eqnarray*}
where, as usual, Card$(\Delta^{-1}(s))$ is the number of
elements of the set $\Delta^{-1}(s)$, and $\Delta$ is the map
\begin{eqnarray*}
\Delta : \mathcal{C} \times \mathcal{C} & \rightarrow 
                    & [0,4R^2] \\
(p,q) & \mapsto & d^2(p,q) \; .
\end{eqnarray*}
Clearly, the distance $d(p,q)$ between sources $p,q \in
\mathcal{C}$ is calculated using the geometry one is concerned
with. The factor 1/2 in the def\/inition of $\eta$ accounts
for the fact that the pairs $(p,q)$ and $(q,p)$ are indeed
the same pair.

The next step is to divide the interval $(0,4R^2]$ in $m$
equal subintervals of length $\delta s = 4R^2/m$. Each
subinterval has the form
\begin{displaymath}
J_i = (s_i - \frac{\delta s}{2} \, , \, s_i + \frac{\delta
s}{2}] \qquad ; \qquad i=1,2, \dots ,m \; ,
\end{displaymath}
with centre
\begin{displaymath}
s_i = \,(i - \frac{1}{2}) \,\, \delta s \;.
\end{displaymath}
The PSH is then obtained from
\begin{equation}
\label{histograma}
\Phi(s_i)=\frac{2}{N(N-1)}\,\,\frac{1}{\delta s}\,
               \sum_{s \in J_i} \eta(s) \;,
\end{equation}
where $N$ is the number of sources in the catalogue $\mathcal{C}$.
The coef\/f\/icient of the sum is a normalization constant such
that
\begin{equation}
\sum_{i=1}^m \Phi(s_i)\,\, \delta s = 1 \, .
\end{equation}
Note that the sum in~(\ref{histograma}) is just a counting of the 
number of pairs of sources separated by a distance whose square 
lies in the subinterval $J_i$, hence $\Phi(s_i)$ is a normalized 
counting. 

It should be stressed that throughout this paper we use normalized 
histograms instead of just plots of countings as in~\cite{LeLaLu} 
and~\cite{Fagundes-Gausmann}. In doing so we can compare histograms 
built up from catalogues with a dif\/ferent number of sources.
Further, although the PSH is actually the plot of the function 
$\Phi(s_i)$, the function $\Phi(s_i)$ itself can be looked upon as 
the PSH. So, in what follows we shall refer to $\Phi(s_i)$ simply 
as the PSH.

As mentioned in the previous section, in a multiply connected
universe the periodic distribution of images on $\widetilde{M}\,$ 
(due to the covering group) gives rise to correlations in their 
positions, and these correlations can be translated into 
correlations in distances between pairs of images. 
For a better understanding on how these distance correlations
arise let us consider the same example used by Fagundes and
Gausmann~\cite{Fagundes-Gausmann} to clarify the method of 
cosmic crystallography. In their work they have assumed that the 
distribution of objects in $M$ is uniform and the catalogue is the 
whole set of observable images. In the Euclidean 3-manifold
they have studied, take a  $g$-pair $(p,gp)$ such that for a 
generic point $p=(x,y,z)$ we have $gp=(x-L,-y+2L,-z)$. The squared 
separation between these points is given by 
\begin{equation}
\label{example}
d^2 = 5L^2 - 8Ly + 4y^2 + 4z^2 \; .
\end{equation}
{}From this equation we have the following: f\/irstly, that the 
separation of any other neighboring $g$-pair%
\footnote{We say that two $g$-pairs $(p,gp)$ and $(q,gq)$ are 
neighbors if the points $p$ and $q$, and thus the points $gp$ and 
$gq$, are neighbors.} 
$(q,gq)$ will be close to $d$; secondly, 
several distant $g$-pairs are separated by approximately 
the same distance; thirdly, one has that not all $g$-pairs
are separated by the same distance (these items hold for the 
isometry $g$ used in this example, of course). Actually, the 
separation of any $g$-pair in Euclidean geometry is independent 
of the pair only when the isometry $g$ is a translation. 
One can sum up by stating that from this example one might 
expect that in general correlations associated to translations 
manifest as spikes in PSH's, whereas correlations due to 
other isometries will be evinced through small 
deviations from the histogram due to uncorrelated pairs.
This conjecture will  be proved in the following two sections.

The distribution of cosmic objects may not be exactly
homogeneous, nor any catalogue will consist of all the observable
sources. For instance, the objects may present some clustering
or may obey a fractal distribution, while luminosity threshold
and obscuration ef\/fects limit our observational
capabilities. We shall show in the following section that 
the consideration of these aspects does not destroy the 
above-described picture, which was qualitatively inferred from 
general arguments and illustrated through the above specif\/ic 
example.

\section{The expected PSH}
\label{pshexp}
\setcounter{equation}{0}

In this section we shall use elements from probability theory
(see for example~\cite{Rohatgi}) to show that the above 
qualitative description of the distance correlations in a PSH
holds in a rather general framework. We shall make clear that 
this picture does not depend on the construction rules one uses
to build a catalogue. Recall that the construction rules to build 
a catalogue $\mathcal{C}$ from an observed universe $\mathcal{U}$
consist of a well-behaved distribution law,
of which the set of objects in $M(t_0)$ is a representative sample,
and of selection rules which dictate how the catalogue is obtained 
{}from the set of all observable images $\mathcal{O}$. 

The general underlying setting of the calculations in this section is 
the existence of an ensemble of catalogues comparable to a given
catalogue $\mathcal{C}$ (real or simulated), with the same number 
of sources $N$ and corresponding to the same constant curvature 
3-manifold $M(t_0)$. So, the construction rules permit the 
computation of probabilities and expected values of quantities which 
depend on the sources in the catalogue $\mathcal{C}$.

Our basic aim now is to compute the expected number, $\eta_{exp}(s_i)$, 
of observed pairs of cosmic sources in a catalogue $\mathcal{C}$ of the 
ensemble with squared separations in $J_i$. Having $\eta_{exp}(s_i)$ we 
clearly have the \emph{expected} pair separation histogram (EPSH) which 
is given by
\begin{equation}
\label{def-EPSH}
\Phi_{exp}(s_i) = \frac{2}{N(N-1)}\,\,\frac{1}{\delta s} \,\,
\eta_{exp}(s_i) \, .
\end{equation}
We remark that the EPSH carries all the relevant information 
of the distance correlations due to the covering group since 
\begin{equation}  \label{noise-def1}
\Phi(s_i) = \Phi_{exp}(s_i) + \mbox{statistical f\/luctuations} \; ,
\end{equation}
where $\Phi(s_i)$ is the PSH constructed with $\mathcal{C}$.

It can be shown (see Appendix A for a proof) that the expected number 
$\eta_{exp}(s_i)$ can be decomposed into its uncorrelated 
part and its correlated part as
\begin{equation}
\label{num-tot}
\eta_{exp}(s_i) = \eta_u(s_i) + \frac{1}{2}\, \sum_{g \in
\widetilde{\Gamma}} \eta_g(s_i) \;,
\end{equation}
where $\eta_u(s_i)$ is the expected number of observed
uncorrelated pairs of sources with squared separations in
$J_i$, i.e. pairs of sources that are not $\Gamma$-pairs;
and $\eta_g(s_i)$ is the expected number of observed $g$-pairs
whose squared separations are in $J_i$. $\widetilde{\Gamma}$
is the covering group $\Gamma$ without the identity map, and
the factor 1/2 in the sum accounts for the fact that, in
considering all non-trivial covering isometries, we are counting
each $\Gamma$-pair twice, since if $(p,q)$ is a $g$-pair,
then $(q,p)$ is a $(g^{-1})$-pair.

For each isometry $g \in \widetilde{\Gamma}$ let us consider
the function
\begin{eqnarray*}
X_g : \widetilde{M}(t_0) & \to & \; [0,\infty) \\
p \quad & \mapsto & d^2(p,gp) \; .
\end{eqnarray*}
This function is a random variable, and using the construction
rules we can calculate the probability of an observed $g$-pair
to be separated by a squared distance that lies in $J_i$,
\begin{equation}
\label{prob-g-pairs}
F_g(s_i) = P[X_g \in J_i] \; .
\end{equation}
The construction rules allow us to compute also the expected
number $N_g$ of $g$-pairs in a catalogue $\mathcal{C}$ with $N$
sources. Clearly in a catalogue with twice the number of sources 
there will be $2N_g$ $g$-pairs. Actually, $N_g$ is proportional 
to $N$ so we write
\begin{equation}   \label{nug}
N_g = N \, \nu_g \; ,
\end{equation}
with $0 \leq \nu_g < 1$.
The expected number of observed $g$-pairs with squared separation
in $J_i$ is thus given by the product of $N_g$ times the
probability that an observed $g$-pair has its squared separation
in $J_i$,
\begin{equation}   \label{nu_g}
\eta_g(s_i) = N \,\nu_g \, F_g(s_i) \; .
\end{equation}

In order to examine uncorrelated pairs, we now consider the
random variable
\begin{eqnarray*}
X : \widetilde{M}(t_0) \times \widetilde{M}(t_0) & \to & [0,\infty) \\
(p,q)\qquad & \mapsto & d^2(p,q) \; .
\end{eqnarray*}
The probability of an observed uncorrelated pair $(p,q)$ to be
separated by a squared distance that lies in $J_i$,
\begin{equation}
F_u(s_i) = P[X \in J_i] \; ,
\end{equation}
can also be calculated from the construction rules. Thus, the
expected number of observed uncorrelated pairs with squared
separation in $J_i$ is
\begin{equation}   \label{eta_u}
\eta_u(s_i) = \left[\,\frac{1}{2}\,N(N-1)- \frac{1}{2}\,N \,\sum_{g
\in \widetilde{\Gamma}} \,\nu_g \,\right] F_u(s_i) \;,
\end{equation}
where we have used~(\ref{nu_g}) and that clearly the expected number 
$N_u$ of uncorrelated pairs is such that
\begin{equation} \label{Nu}
N_u + \frac{1}{2}\,\sum_{g \in \widetilde{\Gamma}} \,N_g =
\frac{1}{2}\, N(N-1) \;.
\end{equation}

{}From~(\ref{def-EPSH}),(\ref{num-tot}), (\ref{eta_u}) and
(\ref{Nu}) an explicit expression for the EPSH is thus
\begin{equation}
\label{EPSH}    
\Phi_{exp}(s_i) = \frac{1}{\delta s}\,\,\frac{1}{N-1}
 \left[\,2\, \frac{N_u}{N} \,\,F_u(s_i) +
\,\sum_{g \in \widetilde{\Gamma}} \,\nu_g\,  F_g(s_i)
                                     \,\right] \; ,
\end{equation}
where the sum in (\ref{EPSH}) is a f\/inite sum since $\nu_g$
is nonzero only for a f\/inite number of isometries. 
For multiply connected manifold $M$  def\/ining 
\begin{eqnarray}
\Phi_{exp}^{u}(s_i) & = & \frac{1}{\delta s}\, F_u(s_i) \;, \label{EPSHu} \\
\Phi_{exp}^{g}(s_i) & = & \frac{1}{\delta s} \,F_g(s_i) \;, \label{EPSHg}
\end{eqnarray}
we obtain a more descriptive expression for the EPSH, namely
\begin{equation}  \label{EPSH2}
\Phi_{exp}(s_i)  = \frac{1}{N-1}\,\, [\,\nu_u\,\Phi_{exp}^{u}(s_i) + 
\, \sum_{g \in \widetilde{\Gamma}} \nu_g\, \Phi_{exp}^g(s_i)
                                 \, ] \;,
\end{equation}
where by analogy with (\ref{nug}) we ave def\/ined 
$\nu_u = 2\,N_u/\,N\,$.
{}From~(\ref{EPSH2}) it is apparent that the EPSH corresponding
to a multiply connected manifold is an EPSH in which the
contributions arising from the $\Gamma$-pairs have been withdrawn, 
plus a term that consists of a sum of individual contributions 
{}from each covering isometry.

When the manifold $M(t_0)$ which gives rise to the comparable 
catalogues is simply connected ($\widetilde{\Gamma}=\emptyset$) 
all $N(N-1)/2$ pairs are uncorrelated, and so from~(\ref{def-EPSH}) 
one clearly has 
\begin{equation} \label{EPSHsc}
\Phi^{sc}_{exp}\,(s_i) = \frac{2}{N(N-1)}\,\frac{1}{\delta s}\,\,
                            \eta^{sc}_{exp}\,(s_i)
                       = \frac{1}{\delta s} \,F_{sc}\,(s_i) \;,
\end{equation}
where $F_{sc}\,(s_i)$ is the probability that the two sources 
be separated by a squared distance that lies in $J_i$. 

It turns out that the combination (which can be obtained
{}from~(\ref{Nu}) and~(\ref{EPSH2}))   
\begin{equation} \label{topsig1}
(N-1)[\,\Phi_{exp}\,(s_i) - \Phi^{sc}_{exp}\,(s_i)\,] =
\nu_u\,\, [\,\Phi^{u}_{exp}\,(s_i) - \Phi^{sc}_{exp}\,(s_i)\,] 
+ \sum_{g \in \widetilde{\Gamma}} \nu_g\,
[\, \Phi^g_{exp}\,(s_i) - \Phi^{sc}_{exp}\,(s_i)\,] 
\end{equation}
proves to exhibit a clear signal of the topology of $M(t_0)$
(see~\cite{GRT00}~--~\cite{GRT01}, which is already available
in gr-qc archive, for more details, including numeric simulation 
and plots). 
This motivates the def\/inition of the following quantity:
\begin{equation}   \label{topsig2}
\varphi^S(s_i) = 
\nu_u\,\,[\,\Phi^{u}_{exp}\,(s_i) - \Phi^{sc}_{exp}\,(s_i)\,] 
      + \sum_{g \in \widetilde{\Gamma}} \nu_g\,
 [\, \Phi^g_{exp}\,(s_i) - \Phi^{sc}_{exp}\,(s_i)\,] \;,
\end{equation}
which throughout this paper is referred to as \emph{topological
signature\/} of the multiply connected manifold $M(t_0)$,
and clearly arises from the ensemble of catalogues of 
discrete sources $\mathcal{C}\,$.
Equation~(\ref{topsig2}) on the one hand makes explicit that 
(i) $\Gamma$-pairs as well as uncorrelated pairs which arise 
(both) from the covering isometries give rise to the topological 
signature; on the other hand, it ensures that (ii) 
the topological signature ought to arise in PSH's even 
when there are only a few images for each object. 

\section{Further results}
\label{fres}
\setcounter{equation}{0}

{}From eqs.~(\ref{prob-g-pairs}) and~(\ref{EPSHg}) we note that 
when $g$ is a Clif\/ford translation (i.e. an isometry such 
that for all $p \in \widetilde{M}(t_0)$, the distance 
$|g|= d(p,gp)$ is independent of $p$) we have
\begin{eqnarray} 
\Phi_{exp}^g(s_i) & = & \left\{ \begin{array}
{l@{\qquad}l}
0 & \mbox{if $\quad |g|^2 \,\notin\, J_i$} \\  (\delta s)^{-1} &
\mbox{if $\quad |g|^2 \,\in \,J_i$ .}
\end{array} \right.
\end{eqnarray}
Thus from equation~(\ref{EPSH2}) one has that the contribution 
of each translation $g$ to the EPSH is a spike of amplitude 
$\nu_g\,[\,(N-1)\delta s\,]^{-1}$ at a well-def\/ined 
subinterval $J_{i_g}$ (say), minus a term proportional to 
the EPSH $\Phi_{exp}^{u}(s_i)$ 
for all $i=1, \dots ,i_g, \dots , m$. On the other hand, 
when $g$ is not a Clif\/ford translation, the separation $|g|$ 
depends smoothly on the $g$-pair because it is a 
composition of two smooth functions: the distance function 
and the isometry $g$. Moreover, the value it takes ranges 
over a fairly wide interval, so $F_g(s_i)$ will be non-zero for 
several subintervals $J_i$. In brief, from~(\ref{EPSH2}) 
we conclude that topological spikes in PSH's are due \emph{only} to 
Clif\/ford translations, whereas other isometries manifest  
as tiny deformations of the PSH corresponding to the 
simply connected case. 

We emphasize that the above general result holds regardless 
of the underlying geometry, and for any set of construction 
rules. Further, when one restricts the above result to specif\/ic
geometries, then one arrives at two rather important consequences,
which we shall discuss in what follows.

Let us f\/irst consider the consequence for the particular case 
of Euclidean manifolds. 
It is known that any compact Euclidean 3-manifold $M$ is 
f\/initely covered by a 3-torus~\cite{Wolf}. Let $\Gamma$ be 
the universal covering group of $M$, then the universal covering 
group of that 3-torus (consisting exclusively of translations) 
is a subgroup of $\Gamma$. Moreover, there is a covering 3-torus 
of $M$ such that its covering group consists of all the translations
contained in $\Gamma$.%
\footnote{The simplest example of this is a cubic 
\emph{half-twisted 3-torus}~\cite{Instantons}, or cubic 
$\mathcal{G}_2$ manifold in Wolf's notation~\cite{Wolf},
which is double covered by a rectangular 3-torus with the same 
square base but with  a height that is the double of the base side. 
This is easily seen by stacking two cubes def\/ining $\mathcal{G}_2$ 
one on top of the other, the common face being the one that is
twisted for an identif\/ication.}  
Thus, PSH's of $M$ and of this minimal covering 3-torus, 
built from comparable catalogues with the same number of sources, 
have the same spike spectrum of topological origin, i.e. the
same set of topological spikes with equal positions and amplitudes.
Therefore the topological spikes alone are not suf\/f\/icient 
for distinguishing these compact f\/lat manifolds, making clear 
that even if the universe is f\/lat ($\Omega_{tot}=1$) the 
spike spectrum is not enough for determining its global 
shape.

Consider now the consequence of our major result for the
special case of hyperbolic manifolds. Since there are no
Clif\/ford translations in hyperbolic geometry~\cite{Ratcliffe},
there are no topological spikes in PSH's built from these manifolds. 
This result was not expected from the outset in that it has been 
claimed that spikes are a characteristic signature of topology 
in cosmic crystallography, at least for f\/lat manifolds%
~\cite{LeLaLu,Fagundes-Gausmann} (see, however, in this regard
the references~\cite{LeLuUz}~--~\cite{FagGaus}).
As a matter of fact, this result is in agreement with simulations
performed in the cases of Weeks~\cite{LeLuUz} and Best~\cite{FagGaus} 
hyperbolic manifolds (see, however, next section for a comparison
of ours and their results).

Incidentally, it can also be f\/igured out from~(\ref{EPSH2}) 
that when $g$ is not a translation, and since the probabilities 
$F_g(s_i)$ are non-zero for several subintervals $J_i$, then 
the contribution of these isometries to the EPSH are in practice 
negligible for $N \approx 2000$ as used in~\cite{LeLaLu} and 
very small for $N \approx 250$ as used in~\cite{Fagundes-Gausmann}. 
In both papers these contributions are hidden by statistical
f\/luctuations and, thus, are not revealed by the isolated
PSH's they have plotted.

{}From what we have seen hitherto it turns out that cosmic 
crystallography, as originally formulated, is not a 
conclusive method for unveiling the shape of our universe 
since (i) in the Euclidean case the topological spikes will 
tell us that spacetime is multiply connected at some scale, 
leaving in some instances its shape undetermined, and (ii) 
in the hyperbolic case, as there are no translations (and 
therefore no topological spikes), it is even impossible to 
distinguish any hyperbolic manifold with non-trivial topology from 
the simply connected manifold $H^3$. Improvements of the cosmic 
crystallography method are therefore necessary. 

\begin{sloppypar}
In the remainder of this section we will brief\/ly discuss a 
f\/irst approach which ref\/ines upon that method.
We look for a means of reducing the statistical f\/luctuations 
well enough to leave a visible signal of the non-translational 
isometries in a PSH. On theoretical ground, the simplest way 
to accomplish this is to use several comparable 
catalogues, with approximately equal number of cosmic sources, 
for the construction of a \emph{mean} pair 
separation histogram (MPSH). For suppose we have $K$ catalogues 
$\mathcal{C}_k$ ($k=1,2,\dots,K$) with PSH's given by
\end{sloppypar}
\begin{equation}  \label{sample-PSH}
\Phi_k(s_i) = \frac{2}{N_k(N_k-1)}\, \frac{1}{\delta s}\,
\sum_{s \in J_i} \eta_k(s)
\end{equation}
with $N_k = \mbox{Card}(\mathcal{C}_k)$. The MPSH def\/ined 
by
\begin{equation} \label{meanPhi}
<\Phi(s_i)>\; = \frac{1}{K} \,\sum_{k=1}^K \Phi_k(s_i)
\end{equation}
is, in the limit $K \to \infty$, approximately equal to
the EPSH. Actually, equality holds when all catalogues have
exactly the same number of sources. However, it can be
shown (see Appendix B for details) that even when the 
catalogues do not have exactly the same number $N$ of
sources, in f\/irst order approximation we still 
have
\begin{equation} \label{limit1}
\Phi_{exp}(s_i) = \lim_{K \rightarrow \infty} <\Phi(s_i)>\;. 
\end{equation}

Elementary statistics tells us that the statistical f\/luctuations 
in the MPSH are reduced by a factor of $1/\sqrt{K}$, which makes
at f\/irst sight the MPSH method very attractive. 
It should be noticed, however, that since in practice it is 
not at all that easy to obtain many comparable real catalogues of 
cosmic sources, this method will hardly be useful in analyses 
which rely on real catalogues. On the other hand, since there is no 
problem in generating hundreds of comparable (simulated) catalogues 
in a computer, the construction of MPSH's can easily be implemented 
in simulations, and so the MPSH method is a suitable approach for 
studying the role of non-translational isometries in PSH's. 
The use of the MPSH technique to extract the
topological signature of non-translational isometries (including 
numeric simulation) is discribed in~\cite{GRT00}~--~\cite{GRT01}.

\section{Conclusions and further remarks}
\label{concl}
\setcounter{equation}{0}

In this section we begin by summarizing our main results,
proceed by brief\/ly indicating possible approaches for further 
investigations, and end by discussing the connection between 
ours and the results recently reported in the referenaces%
~\cite{LeLuUz} and~\cite{FagGaus}.

\vspace{5mm}
\noindent \textbf{\large Main results}
\vspace{3mm}

In this work we have derived the  expression~(\ref{EPSH2})
for the expected pair separation histogram (EPSH) for an ensemble
of comparable catalogues with the same number of sources, and
corresponding to spacetimes whose spacelike sections are any one
of the possible 3-manifolds of constant curvature. The EPSH
is essentially a typical PSH from which the statistical noise 
has been withdrawn, so it carries all the relevant information 
of the distance correlations due to the covering group of $M$. 
The EPSH~(\ref{EPSH2}) we have obtained holds in a rather general 
topological-geometrical-observational setting, that is to say it 
holds when the catalogues of the ensemble obey a well-behaved 
distribution law (needed to ensure that the sources are not 
concentrated in small regions) plus a set of selection rules 
(which dictate how the catalogues ${\cal C\/}$ are obtained 
{}from the set of observable images ${\cal O}\,$). 
It turns out that the EPSH of a multiply connected manifold is 
an EPSH in which the contributions that arise from 
the $\Gamma$-paris have been withdrawn, plus a sum of 
individual contributions from each covering isometry. 
{}From~(\ref{EPSH2}) and (\ref{topsig2}) we have also found 
that the topological signature (contribution of the 
multiply-connectedness) ought to arise in even when there are 
just a few images of each object.

Our theoretical study of distance correlations in pair separation
histograms elucidates the ultimate nature of the spikes and the 
role played by isometries in PSH's. 
Indeed, from the expression~(\ref{EPSH2}) of the EPSH we obtain our 
major consequence, namely that the spikes of topological origin in 
single PSH's are only due to the translations of the covering group, 
whereas correlations due to the other (non-translational) isometries 
manifest as small deformations of the PSH of the underlying 
universal covering manifold. This result holds regardless of 
the (well-behaved) distribution of objects in the universe, and of 
the observational limitations that constrain, for example, the 
deepness and completeness of the catalogues, as long as they 
contain enough $\Gamma$-pairs to yield a clear signal. 

\begin{sloppypar}
Besides clarifying the ultimate origin of spikes and revealing 
the role of non-translational isometries, the above-mentioned major 
result gives rise to two others: 
\end{sloppypar}
\begin{itemize}
\item
That Euclidean distinct manifolds which admit the same 
translations on their covering group present the same 
spike spectrum of topological nature. So, the set of topological
spikes  in the PSH's is not suf\/f\/icient for distinguishing 
these compact f\/lat manifolds, making clear that even if the 
universe is f\/lat ($\Omega_{tot}=1$) the spike spectrum may not 
be enough for determining its global shape;
\item
That individual pair separation histograms corresponding to 
hyperbolic 3-manifolds exhibit no spikes of topological origin, 
since there are no Clif\/ford translations in hyperbolic 
geometry.
\end{itemize}

These two corollaries in turn ensure that cosmic crystallography, 
as originally formulated, is not a conclusive method for unveiling 
the shape of the universe and improvements of the method are
thus necessary.

Any means of reducing the statistical noise well enough for 
revealing the correlations due to non-translational isometries 
should be in principle considered. 
Perhaps the simplest way to accomplish this is through the 
use of the MPSH method that we have also presented; that 
is to say by using several comparable catalogues to construct 
MPSH's. 
The major drawback of this approach, in practice, is the 
dif\/f\/iculty of constructing comparable catalogues of real 
sources. 
Nevertheless, the MPSH method is suitable for studying 
the contributions of non-translational isometries in PSH's by 
computer simulations since there is no problem in constructing 
hundreds of simulated comparable catalogues.
A detailed account of the MPSH technique, including 
numeric simulation, can be found in~\cite{GRT00}~--~\cite{GRT01}.

\vspace{6mm}
\noindent \textbf{\large Further research}
\vspace{3mm}

Any other means of reducing the statistical 
noise may play the role of extracting all distance correlations
instead of just those due to translations. Therefore, a good
approach to this issue is perhaps to study quantitatively the 
noise of PSH's in order to develop f\/ilters. Another scheme for
extracting these correlations from PSH's is to modify what we 
have def\/ined to be an observed universe to make stronger the 
signal which results from the non-translational isometries. 
These ideas are currently under investigation by our research 
group.

The main disadvantages of the known statistical approaches to 
determine the topology of our universe from discrete
sources are that they all assume that: (i) the scale factor
$a(t)$ is accurately known, so one can compute distances
{}from redshif\/ts; (ii) all objects are comoving to a
very good approximation, so multiple images are where
they ought to be; and (iii) the objects have very long
lifetimes, so there exist images of the same object at very 
dif\/ferent distances from one of our images. These are rather
unrealistic assumptions, and no method will be ef\/fective
unless it abandons these simplifying premises. 
Our idea of a suitable choice of data in catalogues (def\/ined 
to be a choice of observed universe) seems to be powerful 
enough to circumvent these problems. Indeed, if one takes as 
observed universe a thin spherical shell, instead of a ball, 
all the sources will be at almost the same distance from the 
observer, and (i) we do not need to know what this distance 
is since we can look for angular correlations between pairs 
of sources, instead of distance correlations, therefore avoiding  
the need of knowing the scale factor, (ii) it is unimportant 
whether they are comoving sources because all of them are now 
contemporaneous, and so (iii) it is also irrelevant if they 
have short lifetimes. In the thin spherical shell one can 
look for angular correlations among $\Gamma$-pairs instead
of the distance correlations of the crystallographic method.
This approach is also currently under study by our research 
group.

\vspace{6mm}
\noindent \textbf{\large A Comparison}
\vspace{3mm}

In what follows we shall discuss the connection between 
our results and those of references~\cite{LeLuUz,FagGaus}. 

The results we have derived regarding PSH's for hyperbolic manifolds 
do not match with the explanation given in~\cite{LeLuUz}
for the absence of spikes. 
Indeed, in~\cite{LeLuUz} it is argued that two types of pairs of 
images can give rise to spikes, namely \emph{type I} and \emph{type II}
pairs.
Considering these types of pairs they argue that in their simulated 
PSH's built for the Weeks manifold there are no spikes because: 
(i) the number of \emph{type I} pairs is too low; and (ii) \emph{type II} 
pairs cannot appear in hyperbolic manifolds.
It should be noticed from the outset that the \emph{type II} pairs 
in~\cite{LeLuUz} are nothing but the $\Gamma$-pairs of the present 
article, whereas \emph{type I} pairs are not the uncorrelated pairs of
this paper. Now, since we have shown that (i) $\Gamma$-pairs as well 
as uncorrelated pairs which arise (both) from the covering isometries 
give rise to the topological signature for multiply-connected 
manifolds, and (ii) the topological signal of $g$-pairs is a spike 
(in a PSH) only if $g$ is a Clif\/ford translation, then the only 
reason for the absence of spikes of topological origin in 
PSH's corresponding to hyperbolic manifolds is that there is no 
Clif\/ford translation in hyperbolic geometry.
Further, that the small number of~\emph{type I} pairs is not 
responsible for the absence of spikes is endorsed by the 
PSH for the Best manifold reported in~\cite{FagGaus}, which was 
performed for an observed universe large enough so that it 
contains in the mean approximately 30 topological images for 
each cosmic object; and yet no spikes whatsoever of topological 
origin were found in the PSH.

Using expression~(\ref{EPSH2}) for the EPSH  one can also 
clarify the ef\/fect of subtracting from the
PSH corresponding to a particular 3-manifold the PSH of
the underlying simply connected covering manifold. This type
of dif\/ference has been performed for simulated 
comparable catalogues
(with equal number of images and identical cosmological 
parameters) for a Best and $H^3$ manifolds~\cite{FagGaus}.
In general the plots of that dif\/ference exhibit a fraction 
$1/(N-1)$ of the topological signature of the isometries plus 
(algebraically) the f\/luctuations corresponding to both PSH's 
involved, namely the PSH for the underlying simply connected 
space and the PSH for the multiply connected 3-manifold itself. 
To understand that this is so, let us rewrite eq.~(\ref{noise-def1}) 
as 
\begin{equation}  
\label{noise-def2}
\Phi(s_i) = \Phi_{exp}(s_i) + \rho\,(s_i) \; ,
\end{equation}
where $\rho\,(s_i)$ denotes the noise (statistical f\/luctuations) of 
$\Phi(s_i)$.
Using  the decomposition~(\ref{noise-def2}) together with~%
(\ref{topsig2}) one easily obtains
\begin{equation}   \label{FG-dif}
\Phi(s_i) - \Phi^{sc}(s_i) = \frac{1}{N-1}\,\varphi^{S}(s_i) 
                       + \, \rho\,(s_i)- \rho^{sc}(s_i) \; ,
\end{equation}
where $\varphi^{S}(s_i)$ denotes the topological signature, 
which is given by~(\ref{topsig2}).
According to (\ref{topsig1}), (\ref{topsig2}) and (\ref{noise-def2}), 
had they examined the dif\/ference $\Phi(s_i) - \Phi_{exp}^{sc}(s_i)$, 
between the PSH built from that Best manifold and the EPSH for the 
corresponding  covering space, they would have expurgated 
the noise $\rho^{sc}(s_i)$, and thus their plot 
(f\/ig.~3 in~\cite{FagGaus}) would simply contain 
a superposition of the topological signature 
and just one noise,  $\rho\,(s_i)$. 
As a matter of fact, the ``wild oscillations in the scale of the 
bin width'' they have found are caused by the superposition of 
the two statistical f\/luctuations $\rho^{sc}(s_i)$ and 
$\rho\,(s_i)$, whereas the ``broad pattern on the scale of 
$R_0$'' ought to carry basic features of the topological 
signature corresponding to  the Best manifold they have examined. 
But again, a detailed account of these points is a matter that
has been discussed in~\cite{GRT00}~--~\cite{GRT01}.

Finally, regarding the peaks of ref.~\cite{Fagundes-Gausmann} 
it is important to bear in mind that graphs tend to be ef\/fective 
mainly to improve the degree of intuition, to raise questions to 
be eventually explained, and for substantiating theoretical results. 
Often, however, they do not constitute a proof for a result, such 
as the EPSH~(4.12) we have formally derived from rather general 
f\/irst principles. A good example which shows the limitation 
of the conclusions one can withdraw from such graphs comes exactly
{}from the PSH shown in f\/ig.~1 of ref.~\cite{Fagundes-Gausmann}, 
where there is a signif\/icant peak which according to 
our results clearly {\em cannot\/} be of topological origin
because it does not correspond to any translation. 
Note, however, that just by examining that graph one cannot at all 
decide whether that sharp peak is of topological nature or 
arises from purely statistical f\/luctuations. 
In brief, just by examining PSH's one cannot at all distinguish between 
spikes of topological origin from sharp peaks of purely statistical 
nature --- our statistical analysis of the distance correlations in 
PSH's elucidates the ultimate role of all types of isometries (in PSH's) 
and are necessary to separate the spikes (sharp peaks) of 
dif\/ferent nature.

\section*{Acknowledgements}

We thank the scientif\/ic agencies CNPq, CAPES
and FAPERJ for f\/inancial support. G.I.G. is grateful
to Helio Fagundes for introducing him to the method of
cosmic crystallography and to Evelise Gausmann for fruitful
conversations.

\appendix
\section*{Appendix A}
\renewcommand{\theequation}{A.\arabic{equation}}

Our aim in this appendix is to show that the $\eta_{exp}(s_i)$ 
can be decomposed into an uncorrelated part and a correlated
part according to~(\ref{num-tot}). To this end suppose
that there are  $K$ comparable catalogues $\mathcal{C}_k$, 
each of which contains the same number $N$
of sources and corresponds to the constant curvature 
3-manifold manifold $M(t_0)$. Since any pair of sources in each 
catalogue is either a $\Gamma$-pair or an uncorrelated pair
(a pair that is not a $\Gamma$-pair) the number $\eta^{(k)}(s_i)$ 
of pairs of sources in a catalogue $\mathcal{C}_k$ with squared 
separations in $J_i$ splits as
\begin{equation}
\label{num-tot1}
\eta^{(k)}(s_i) = \eta_u^{(k)}(s_i) + \frac{1}{2}\, \sum_{g \in
\widetilde{\Gamma}} \eta_g^{(k)}(s_i) \;,
\end{equation}
where $\eta_u^{(k)}(s_i)$ is the number of uncorrelated pairs of 
sources in $\mathcal{C}_k$ with squared separations in $J_i$, and 
$\eta_g^{(k)}(s_i)$ is the number of $g$-pairs in $\mathcal{C}_k$ 
whose squared separations are in $J_i$. $\widetilde{\Gamma}$ is 
the covering group $\Gamma$ without the identity map, and the 
factor 1/2 in the sum accounts for the fact that, in considering 
all non-trivial covering isometries, we are counting each 
$\Gamma$-pair twice, since if $(p,q)$ is a $g$-pair, then $(q,p)$ 
is a $(g^{-1})$-pair.

Taking the mean value of $\eta^{(k)}(s_i)$ in the set of $K$ 
catalogues we have
\begin{equation}
\label{mean-num-tot}
<\eta(s_i)>\: = \; <\eta_u(s_i)> + \, \frac{1}{2}\, \sum_{g \in
\widetilde{\Gamma}} <\eta_g(s_i)> \;,
\end{equation}
where
\begin{equation}
\label{mean-num}
<\eta(s_i)>\; = \: \frac{1}{K} \sum_{k=1}^K \eta^{(k)}(s_i) \;,
\end{equation}
and analogous expressions hold for $\,<\eta_u(s_i)\,>$ and 
$\,<\eta_g(s_i)>\,$. 
Now, since  expected values the limit of their corresponding
mean values when the number of samples $K$ tends to 
inf\/inity, one has
\begin{equation}
\eta_{exp}(s_i) = \lim_{K \rightarrow \infty} <\eta(s_i)> \;,
\end{equation}
and similar expressions hold for $<\eta_u(s_i)>$ and 
$<\eta_g(s_i)>$. Now, using~(\ref{mean-num-tot}) we obtain 
\begin{equation}
\label{exp-num-tot}
\eta_{exp}(s_i) = \eta_u(s_i) + \frac{1}{2}\, \sum_{g \in
\widetilde{\Gamma}} \eta_g(s_i) \;,
\end{equation}
where $\eta_u(s_i)$ is the expected number of observed
uncorrelated pairs of sources with squared separations in
$J_i$, and $\eta_g(s_i)$ is the expected number of observed 
$g$-pairs whose squared separations are in $J_i$.

\section*{Appendix B}
\renewcommand{\theequation}{B.\arabic{equation}}

Our aim in this appendix is to show that equation~(\ref{limit1})
holds in f\/irst approximation. To this end let us write
\begin{equation}
N_k = N + \Delta N_k \;,
\end{equation}
for $k=1, \dots ,K$. Since we are assuming that $\Delta N_k$ is 
small ($\Delta N_k \ll N$), in f\/irst order approximation in 
$\Delta N_k$ the normalization coef\/f\/icient of
(\ref{sample-PSH}) can clearly be expanded to give 
\begin{equation}
\frac{1}{N_k(N_k-1)} \,\approx \, \frac{1}{N(N-1)} 
             - \frac{2N-1}{[N(N-1)]^2} \,\,\Delta N_k \; ,
\end{equation}
so that the mean value of $\Phi(s_i)$ in this approximation
reduces to 
\begin{equation}
\label{mean-Phi1}
<\Phi(s_i)> \; \approx \; \frac{2}{N(N-1)} \,\frac{1}{\delta s} 
  \left[\,<\eta(s_i)> -\,\frac{2N-1}{N(N-1)}\, <\Delta N\, \eta(s_i)>
                \right] \;,
\end{equation}
where 
\begin{equation}
<\Delta N \,\eta(s_i)> \: = \,\frac{1}{K}\, \sum_{k=1}^K  
                  \Delta N_k \,\,\eta^{(k)}(s_i) \;.
\end{equation}
However, since the quantities $\Delta N_k$ and $\,\eta^{(k)}(s_i)$ 
are statistically independent, then the equation
\begin{equation} \label{indep}
<\Delta N \,\eta(s_i)>\: = \;<\Delta N> \,\, <\eta(s_i)> %\;.
\end{equation}
holds. Now, inserting this equation into~(\ref{mean-Phi1}) 
one obtains
\begin{equation}
\label{mean-Phi2}
<\Phi(s_i)>\: \approx \; \left[\,1 - \frac{2N-1}{N(N-1)}<\Delta N> \right] 
\,\,  \frac{2}{N(N-1)}\, \frac{1}{\delta s}\, <\eta(s_i)> \;.
\end{equation}
In the limit $K \rightarrow \infty$ we have that $<\Delta N>\, \rightarrow 0$ 
and $<\eta(s_i)>\, \rightarrow \eta_{exp}(s_i)$. Thus, in f\/irst order 
approximation we have
\begin{equation}
\label{limit}
\lim_{K \rightarrow \infty} <\Phi(s_i)>\; \approx\; \Phi_{exp}(s_i) \;,
\end{equation}
which completes the proof.

\end{document}